\documentclass[a4paper,11pt]{article}
\usepackage{jinstpub} 
\usepackage{lineno}

\usepackage{multirow,booktabs}

\title{\boldmath A new hydrogen-filled Cherenkov detector for Kaon tagging at the NA62 experiment at CERN}

\author[a,1]{Jack Sanders\note{On behalf of the NA62 Collaboration}}
\affiliation[a]{The University of Birmingham,\\
Birmingham, UK}

\emailAdd{jack.sanders@cern.ch}

\abstract{The precision measurement of $K^{+} \rightarrow \pi^{+} \nu \bar{\nu}$ at the NA62 experiment requires a kaon identification detector to have a time resolution better than 100\,ps, at least 95\,\% kaon identification efficiency, and a pion misidentiﬁcation probability of less than $10^{-4}$. Since the start of NA62 data taking in 2016, kaon identification has been performed by a differential Cherenkov with achromatic ring focus (CEDAR) detector with a nitrogen gas radiator. A new CEDAR using hydrogen (CEDAR-H) as a radiator gas has been developed to reduce the material in the beamline, reducing the beam particle scattering within the detector. 
CEDAR-H was validated during a two-week test beam at CERN in 2022 and was approved by the NA62 collaboration for use in data taking from 2023. The test beam results, installation and commissioning in the NA62 beamline are reported. }

\keywords{Cherenkov and transition radiation, Timing detectors }

\begin{document}
\maketitle

\flushbottom

\section{Introduction}\label{sec:intro}
The NA62 experiment \cite{Gil_2017} is a fixed target experiment located in the north area of CERN specifically designed to measure the branching fraction of the ultra-rare charged kaon decay $K^+ \rightarrow \pi^+ \nu \bar{\nu}$. NA62 is provided with a 400\,GeV$/c$ proton primary beam from the CERN SPS, incident on a Be target, resulting in an unseparated secondary hadron beam. The secondary hadron beam has a mean momentum of 75\,GeV$/c$, with a composition of pions (70\,\%), protons (23\,\%) and kaons (6\,\%) with a nominal beam particle rate of 600\,MHz. It is essential that the kaons are selected from the unseparated hadron beam to suppress backgrounds caused by interactions of beam pions with material in the beamline. Strict requirements for a kaon identification detector are in place: a timing resolution of $<100$\,ps, with a kaon identification efficiency $>95\,\%$ and a pion misidentification probability $<10^{-4}$. 

\section{The NA62 detector}\label{sec:NA62Exp}
\begin{figure}[htbp]
\centering
\includegraphics[width=.9\textwidth]{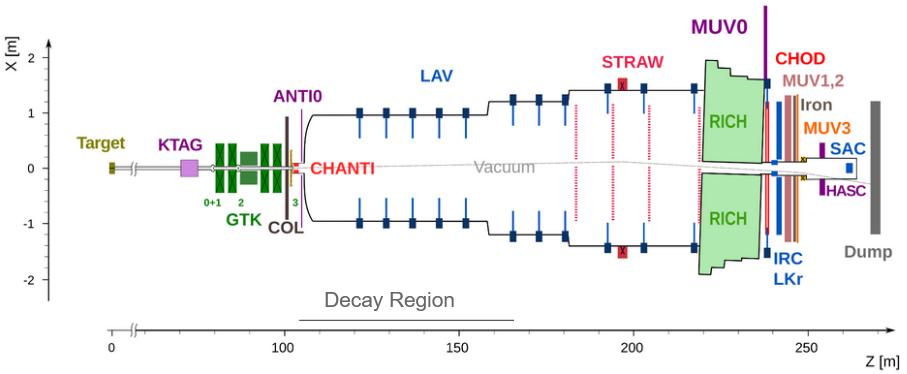}
\caption{X-Z view of the NA62 beamline and detector in 2021.\label{fig:na62det}}
\end{figure}

The NA62 detector setup can be found in Figure \ref{fig:na62det}. Charged kaons are selected from the unseparated hadron beam via a differential Cherenkov with achromatic ring focus (CEDAR) detector combined with a purpose-built photon detection system (KTAG) located 70\,m downstream of the target. The KTAG provides a precise kaon time reference for event reconstruction. The beam particles' momenta and directions are measured by a beam spectrometer (GTK), which comprises four silicon pixel detectors located 80-102\,m from the target. The hadron beam then passes through a 65\,m long decay region, where 10\,\% of the kaons decay. The STRAW spectrometer is found downstream of the decay region, consisting of four chambers situated between 180 and 220\,m from the target. When combined with the MNP33 dipole magnet, the STRAW measures the trajectories and momenta of the charged products of $K^+$ decays. 
A particle identification system is employed, comprised of a 17\,m long ring-imaging Cherenkov counter (RICH) designed to separate final state $\pi^+$ and $\mu^+$ and a calorimetry system comprised of an electromagnetic calorimeter (LKr) and hadronic calorimeters (MUV1,2). A photon veto system (12 LAV stations, LKr, IRC, SAC) provides 0-50\,mrad photon veto coverage.
\section{The CEDAR detector}\label{sec:CEDAR}
\begin{figure}[htbp]
\centering
\includegraphics[width=.9\textwidth]{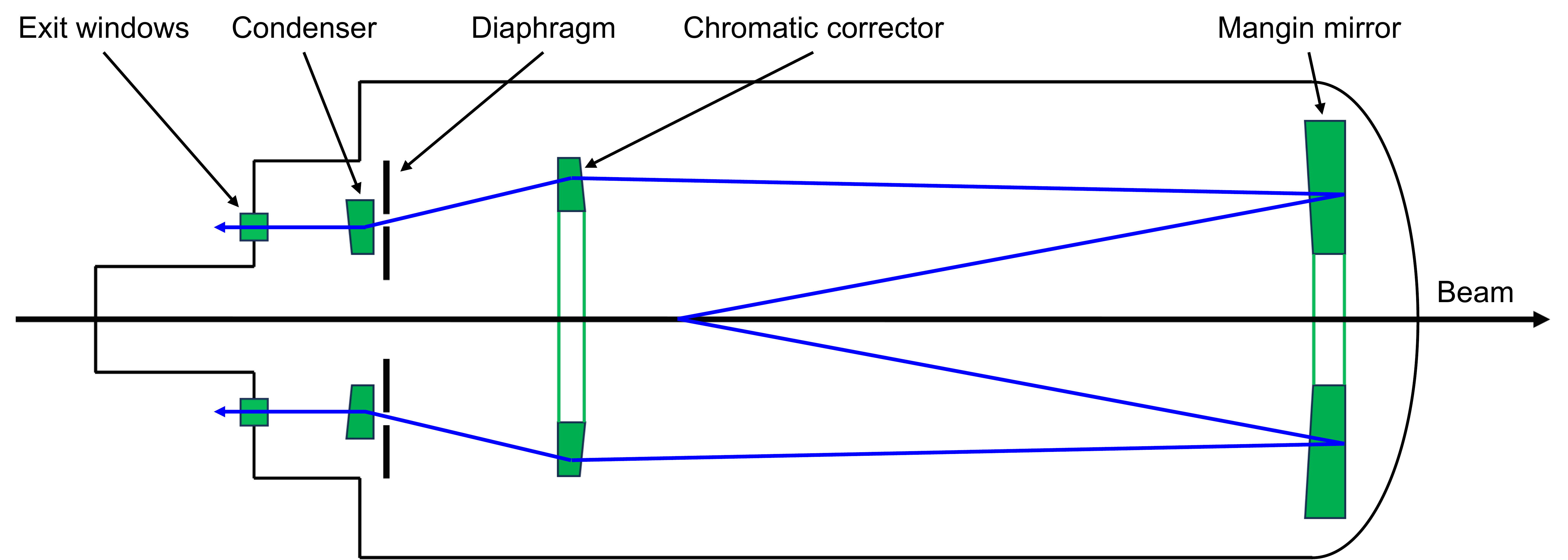}
\caption{Sketch of a CEDAR with the vessel outer wall (black lines), the internal optics (green areas) and an example Cherenkov light path (blue lines).\label{fig:CedarDia}}
\end{figure}

CEDARs were developed at CERN \cite{Bovet:1982xf} for secondary beam particle detection at the super proton synchrotron (SPS). A sketch of the CEDAR detector can be found in Figure \ref{fig:CedarDia}. A CEDAR comprises a thermally insulated gas vessel with a volume of $1.1\,\mbox{m}^3$ consisting of two connected sections: a main cylindrical vessel of length 4.5\,m with an inner diameter of 55\,cm; and a smaller diameter vessel end cap of either 28\,cm or 34\,cm in length, dependent on the CEDAR type. 
The smaller diameter vessel allows for the placement of eight quartz exit windows close to the beam axis.
The vessel is connected to a beam pipe at either end, with aluminium windows separating the gas from the vacuum in the beam pipes. 

The CEDAR optics are contained within the gas vessel and are aligned along the longitudinal axis, shown in Figure \ref{fig:CedarDia}. The optical system is designed for a particle beam with a divergence of less than $100\mu$m, so that Cherenkov photons lie on concentric cones. Cherenkov light produced by beam particles in the gas radiator is first reflected by a Mangin mirror \cite{Riedl:74} located at the downstream end of the detector. The mirror comprises a lens with two spherical surfaces, with the downstream surface cemented onto a concave mirror designed to reduce any spherical aberrations. The light then passes through a Chromatic corrector, a plano-convex lens that is designed to correct for chromatic dispersion within the gas and the lenses. The combination of the optical elements focuses the light into a ring at the position of the diaphragm, which has a variable annular aperture centred at 100\,mm, allowing for aperture widths between 0 and 20\,mm. Condenser lenses focus the Cherenkov light that passes through the diaphragm aperture, and the light then exits the CEDAR vessel through the eight quartz windows.

The radius of the Cherenkov ring depends on the radiator gas density inside the CEDAR vessel. The gas pressure can be varied to obtain a Cherenkov ring at a radius of 100\,mm at the position of the diaphragm for the selected beam particle. The NA62 CEDAR used nitrogen (N$_2$) as a radiator gas. At the nominal hadron beam momentum of 75GeV$/c$, a pressure of 1.71\,bar at a temperature of 293\,K produced a Cherenkov ring with a mean radius of 100\,mm at the diaphragm for a charged kaon. 
At this pressure, the pion and proton rings had radii of 102\,mm and 94\,mm respectively. A diaphragm aperture of 1.3\,mm was used to separate the light from the kaon and the other beam components.

In the original CERN design, the Cherenkov light exiting each CEDAR exit window was detected by an ET-9820QB photomultiplier tube (PMT) placed directly on the exit window. A particle was identified if there was a coincidence of at least six PMTs.
In NA62, the light from the quartz exit windows is reflected radially by eight spherical mirrors onto 8 PMT planes, referred to as sectors. The PMT planes are equipped with an array of 48 single-anode Hamamatsu PMTs of two types: 32 R9880-110 and 16 R7400. The KTAG photo-detection system allowed the use of at least five sectors for particle identification, resulting in an improved kaon identification efficiency compared to the original setup.
At NA62, a kaon time resolution of $70$\,ps and a kaon identification efficiency of $95\%$ was achieved with the KTAG and CEDAR \cite{GOUDZOVSKI201586}.

\section{CEDAR-H development}
For the precise measurement of ultra-rare kaon decays, it is essential to reduce or identify backgrounds from the interaction of beam particles with the material in the beamline. The impact of these backgrounds can be reduced by detecting scattered beam particles or secondary particles before the decay regions using vetoing detectors or by reducing the material in the beamline. 
The NA62 CEDAR with N$_{2}$ radiator gas at 1.71\,bar contributes the most significant amount of material in the beamline. The material in the path of the beam due to the NA62 CEDAR is $39 \times 10^{-3}\,X_{0}$, with the N$_{2}$ radiator gas comprising $35 \times 10^{-3}\,X_{0}$ and the aluminium windows $3.9 \times 10^{-3}\,X_{0}$, where $X_{0}$ is one radiation length. 

A CEDAR with hydrogen (H$_{2}$) as a radiator gas at a pressure of 3.85\,bar produces a similar Cherenkov angle to N$_2$ at 1.71\,bar. Using H$_{2}$ reduces the material in the beamline to $7.3 \times 10^{-3}\,X_0$, with $3.4 \times 10^{-3}\,X_0$ from the H$_{2}$ radiator gas and the remaining contribution from the aluminium windows.
This corresponds to a significant reduction of the material in the beamline. However, simply replacing the radiator gas in the NA62 CEDAR from N$_{2}$ to H$_{2}$ would not have produced a satisfactory kaon tagging performance. 
The loss in performance is because the optics of the NA62 CEDAR do not correctly account for the chromatic dispersion in H2.
As no suitable CEDAR existed, a new detector was developed by refitting an existing CEDAR vessel with new optics designed for use with H$_{2}$ (CEDAR-H).

A detailed simulation in GEANT4 \cite{GEANT4:2002zbu} of the CEDAR, its internal optics, and the KTAG was used to design new optics for use with H$_{2}$. An iterative procedure of modifying the optical parameters of the Chromatic corrector and the Mangin mirror was used to maximise the number of photons reaching the KTAG PMT arrays. 
The results of the iterative design process produced significant modifications to the radii of curvature of the Chromatic corrector and Mangin mirror. The inner radius of the Mangin mirror was also reduced to increase the overall reflective surface. The radius of curvature of the spherical mirrors was modified to improve the distribution of light on the PMT arrays in the KTAG. After the design process, the CEDAR-H was constructed at CERN in 2022. 
\section{CEDAR-H test beam at CERN}
\begin{figure}[htbp]
\centering
\includegraphics[width=.75\textwidth]{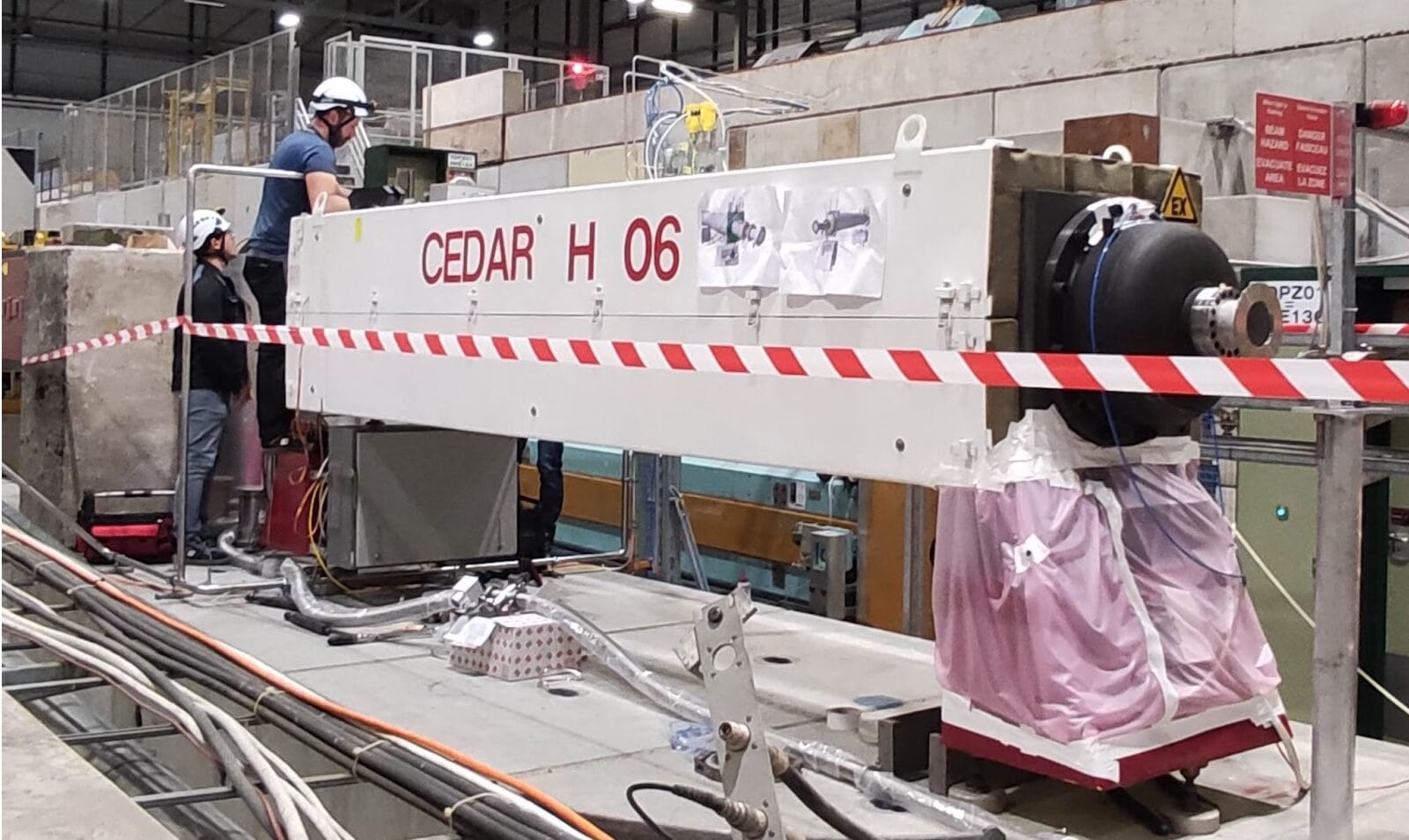}
\caption{CEDAR-H installed at the H6 beamline at the CERN SPS .\label{fig:bealineInstall}}
\end{figure}
CEDAR-H was installed into the H6 beamline 440\,m downstream of the target at the CERN SPS in October 2022 (Figure \ref{fig:bealineInstall}). A primary 400\,GeV$/c$ proton beam, provided by the SPS, incident on the target produced an unseparated secondary hadron beam similar to the NA62 beam, with a mean beam momentum of 75\,GeV$/c$. The difference in distance from the target resulted in a different beam composition compared to NA62: pions (71\,\%), protons (25\,\%) and kaons (4\,\%).

The test beam aimed to validate the internal optical components of CEDAR-H and to measure the kaon photon yield and the pion misidentification probability. CEDAR-H was equipped with eight ET-9280QB PMTs, with one placed on each of the eight quartz exit windows. A pair of scintillating chambers were placed at the upstream and downstream ends of the CEDAR-H vessel and were used to provide a trigger signal for the beam particles passing through the detector. 

\begin{figure}[htbp]
\centering
\includegraphics[width=.5\textwidth]{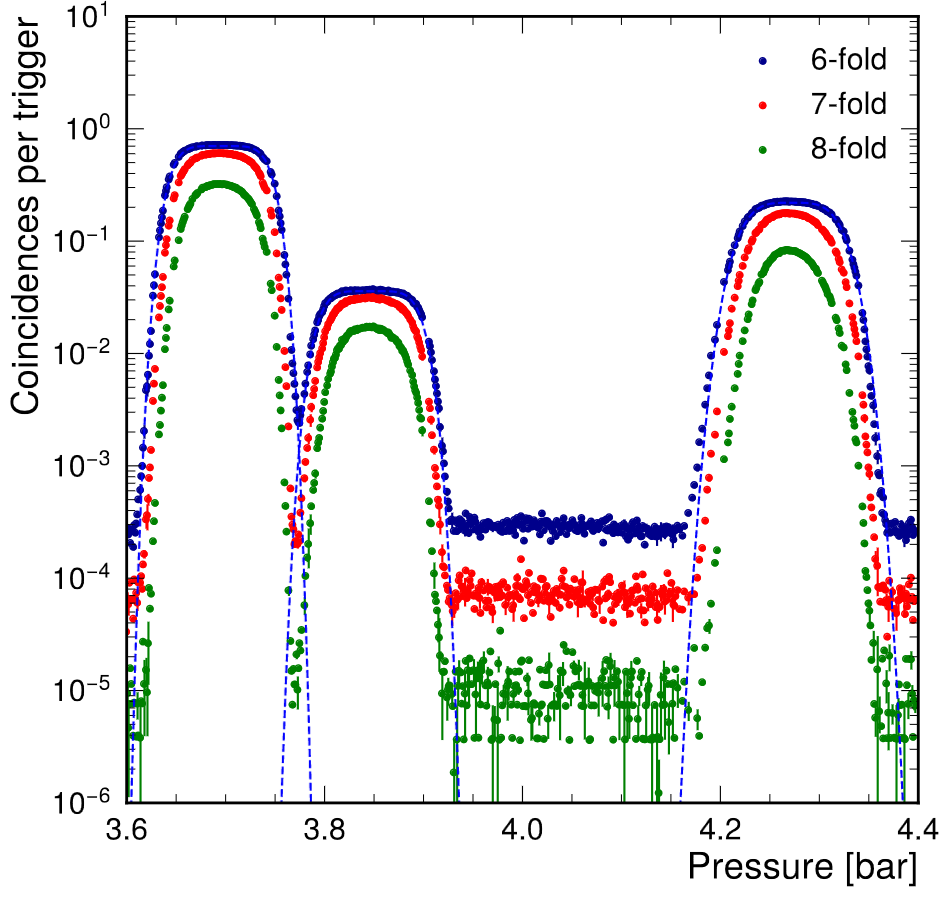}
\caption{Pressure scan results for CEDAR-H from the H6 test beam with a diaphragm aperture of 1.7\,mm, showing the number of coincidences inclusive. The left peak shows coincidences produced by the pion, the centre peak from the kaon and the right peak from the proton. The 6-fold coincidence peaks are fitted, shown by blue dashed lines, to compute the pion misidentification probability. \label{fig:testbeamres}}
\end{figure}

Several pressure scans were completed at diaphragm apertures between 1.3 and 2.3\,mm. During a pressure scan, the pressure was varied between 3.6 and 4.4\,bar, and the PMT coincidences were recorded. The change in pressure allowed the CEDAR to be sensitive to the different beam components. After several pressure scans, an optimal diaphragm aperture of 1.7\,mm was selected. The results of the pressure scan are reported in Figure \ref{fig:testbeamres}. The pressure at which the kaon beam particle produced the maximum coincidences was at 3.85\,bar, which corresponded to a light yield of 19.1 photo-electrons. 
The pion misidentification probability was estimated from a fit to the data from events with at least 6-fold coincidence at the kaon pressure, shown by the blue dashed lines in Figure \ref{fig:testbeamres}. At a pressure of 3.85\,bar, the pion misidentification probability was measured to be less than $10^{-4}$. 
\section{CEDAR-H commissioning at NA62}
\begin{figure}[htbp]
\centering
\includegraphics[width=.75\textwidth]{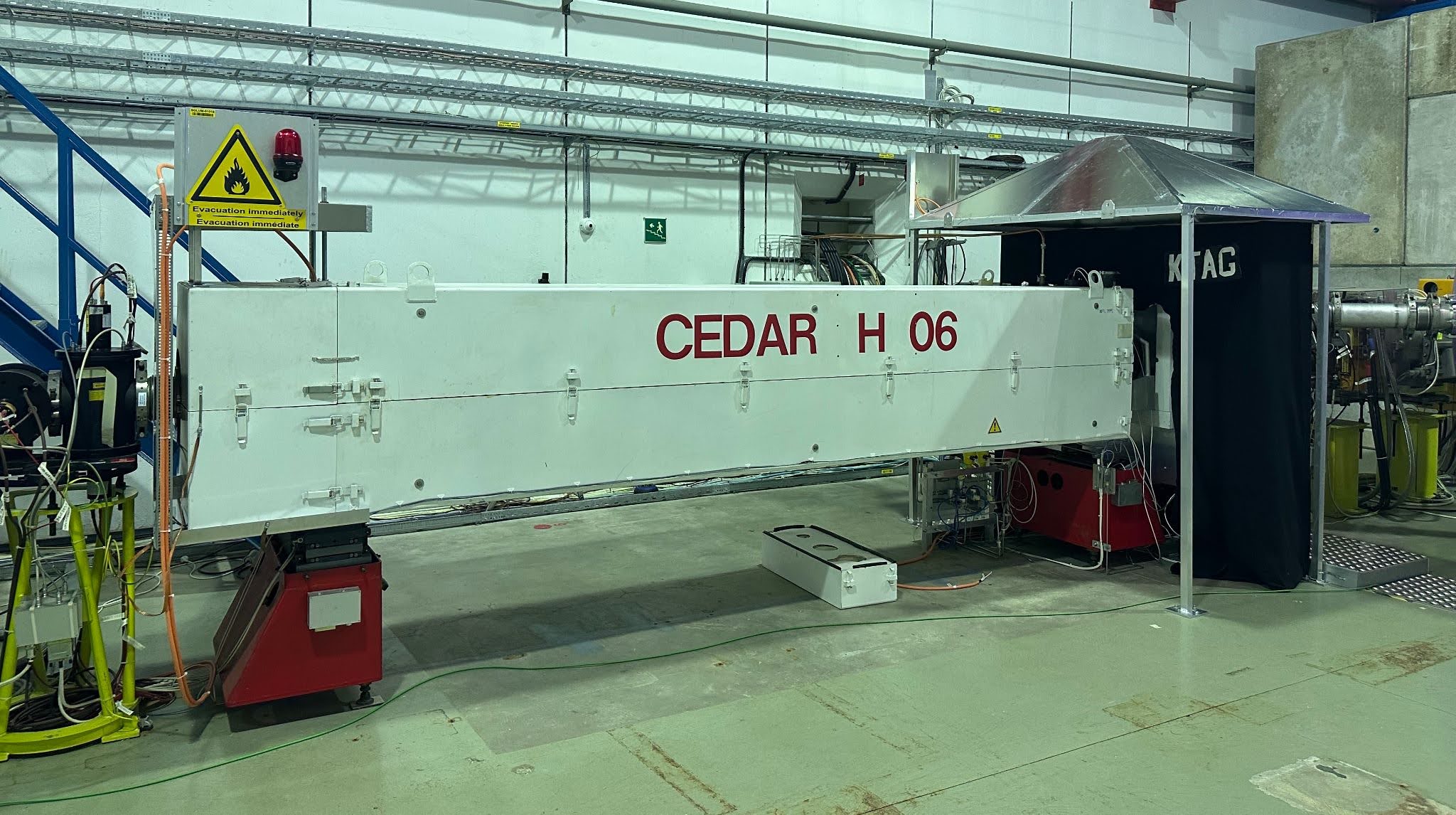}
\caption{CEDAR-H installed in the NA62 beamline, with the flammable gas safety hoods on both the downstream end (left) and the upstream end above the black KTAG enclosure (right). The XY motors are found in the red foot at the downstream end (left). 
\label{fig:na62install}}
\end{figure}

CEDAR-H was installed on the NA62 beamline at the beginning of 2023. The KTAG was disassembled from the NA62 CEDAR, and the CEDAR was removed from the beamline. CEDAR-H was transported to the NA62 experimental hall, where it was installed on the beamline, and the KTAG was reassembled and mounted to the upstream end of the detector (Figure \ref{fig:na62install}). A number of safety systems were installed, including flammable gas detectors found at the top of the silver hoods in Figure \ref{fig:na62install}. 

At the start of data-taking at NA62 in 2023, the CEDAR-H vessel was aligned with the beam in order to maximise the number of Cherenkov photons passing through the diaphragm from a beam kaon. Using the XY motors located in the downstream foot (Figure \ref{fig:na62install}) of the CEDAR-H and the fixed upstream end, a fine angular detector alignment was performed at the kaon pressure, where the number of photo-electrons per kaon was maximised.  

\begin{figure}[htbp]
\centering
\includegraphics[width=.45\textwidth]{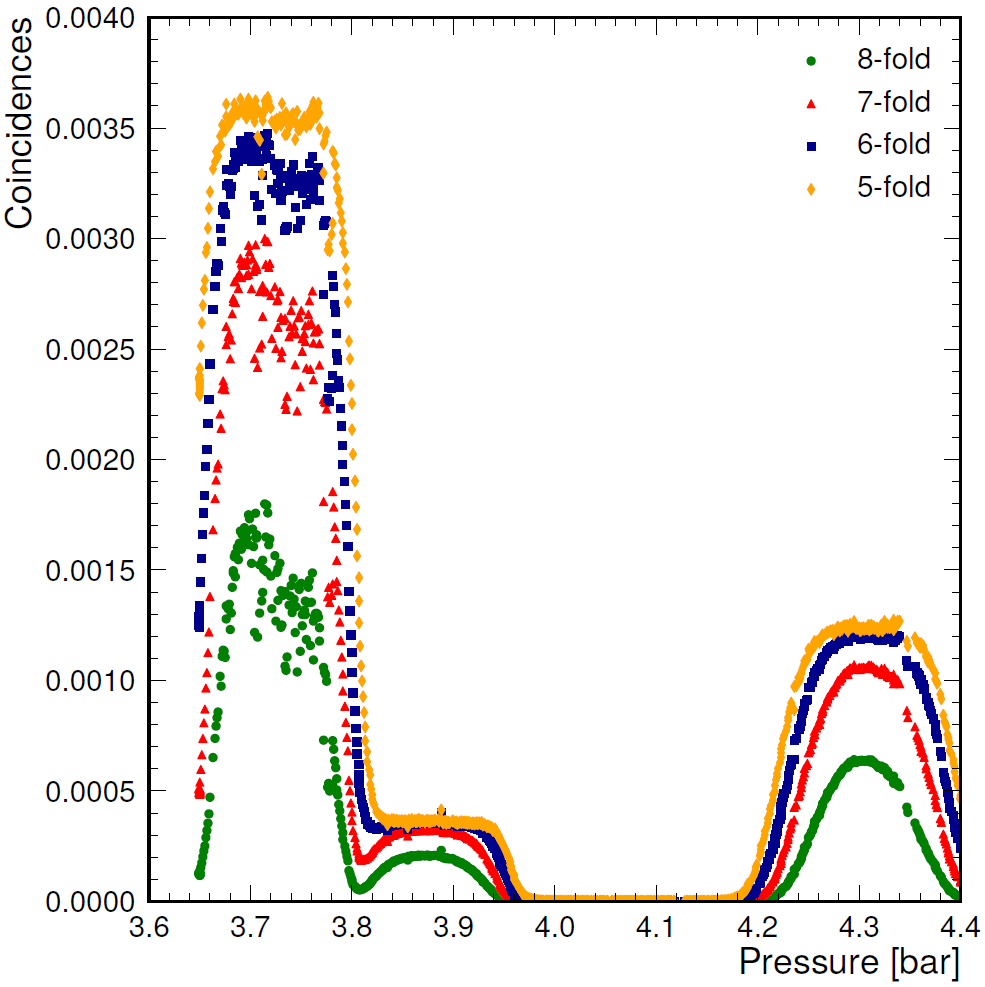}
\qquad
\includegraphics[width=.45\textwidth]{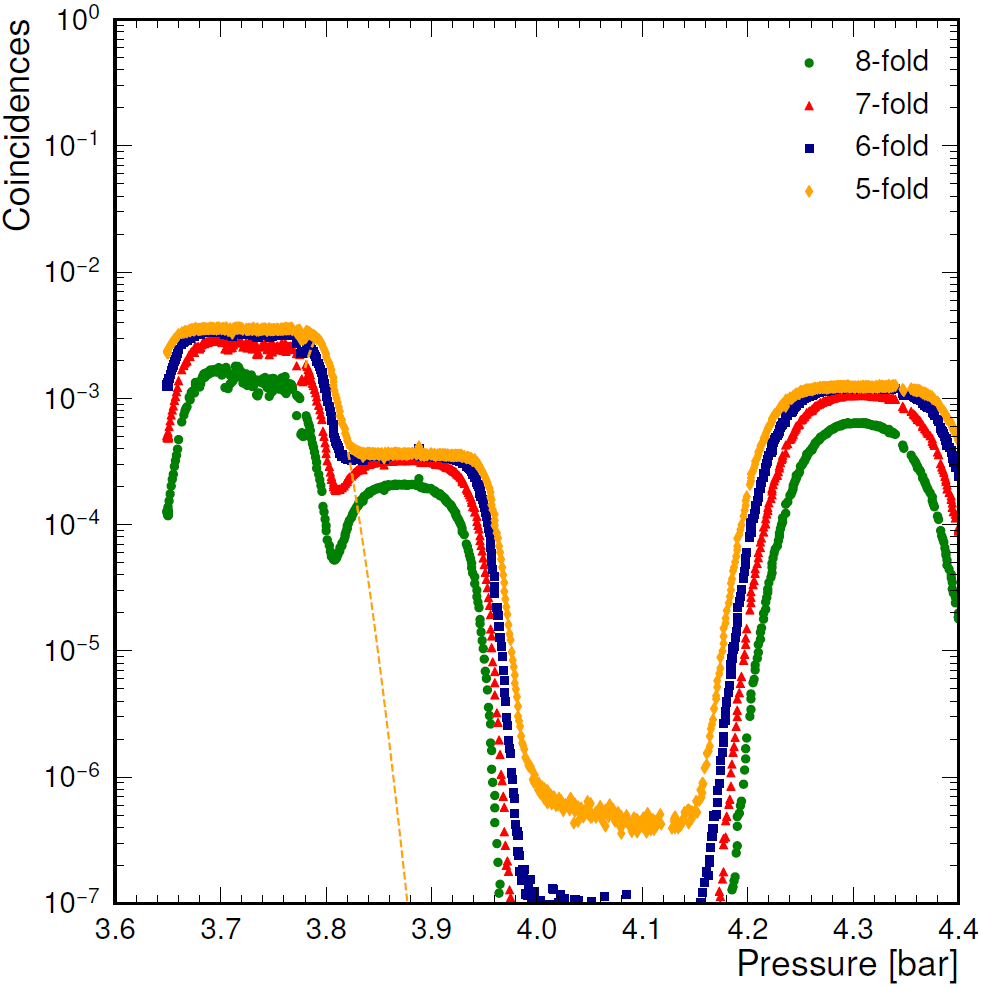}
\caption{Number of coincidences (normalised by beam intensity) inclusive at different pressures with a diaphragm aperture of 1.8\,mm. The left panel is in linear scale, and the right panel is in log scale. The 5-fold coincidences for the pion are fitted in the right panel (orange dashed line) to estimate the pion light contribution at the kaon pressure of 3.88\,bar. \label{fig:Na62Res}}
\end{figure}

As in the test beam, several pressure scans were completed at different diaphragm apertures. A diaphragm aperture of 1.8\,mm was selected for its excellent light yield and low pion misidentification probability. The results of the pressure scan at 1.8\, mm are shown in Figure \ref{fig:Na62Res}. At the kaon pressure of 3.88\,bar, a light yield of 21 photo-electrons per kaon was measured with the pion misidentification probability estimated as less than $10^{-4}$ using a fit to the five-fold coincidence in Figure \ref{fig:Na62Res}.

\begin{figure}[htbp]
\centering
\includegraphics[width=.45\textwidth]{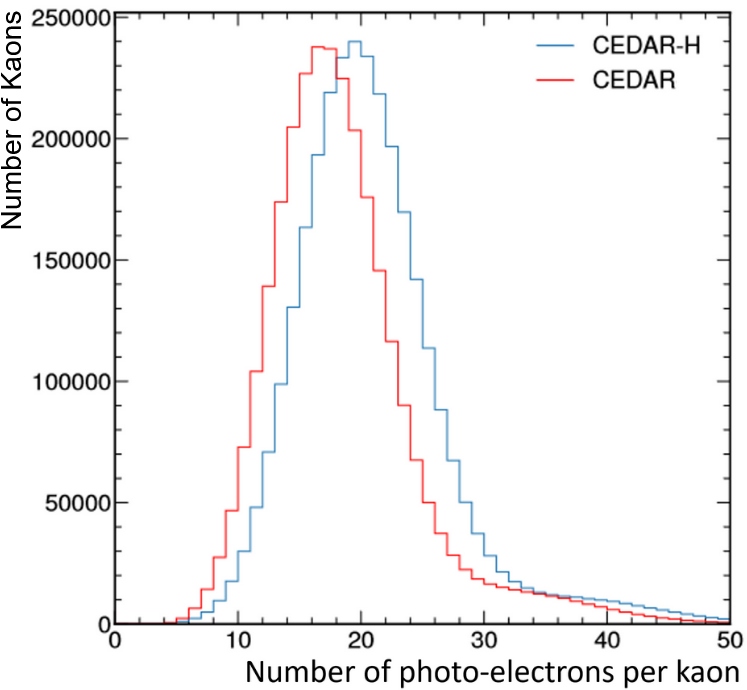}
\caption{Number of photo-electrons per kaon for the NA62 CEDAR (red) and CEDAR-H (blue) \label{fig:Na62comp}}
\end{figure}
\section{Summary}
The NA62 experiment has installed a new CEDAR detector designed to reduce the material in the beam's path and, therefore, reduce the scattering of beam particles. This was completed by switching the radiator gas from N$_2$ at 1.71\,bar to H$_2$ at 3.85\,bar.
The change in radiator gas required a redesign of the internal optics of the CEDAR, which was completed at the University of Birmingham using a full GEANT4 simulation and was then constructed at CERN in 2022. 
CEDAR-H was validated during a CERN test beam and was installed on the NA62 beamline at the beginning of 2023. CEDAR-H was commissioned at the start of the NA62 2023 run, where the detector was aligned, and several pressure scans were completed. 
The gas pressure corresponding to a maximum light yield from kaons was found to be 3.88\,bar, and the optimal diaphragm aperture allowing for the collection of light from kaons and satisfactory pion-kaon separation was found to be 1.8\,mm. 
With this setting, a kaon beam particle produced a light yield of 21 photo-electrons, with a pion misidentification probability of less than $10^{-4}$.

CEDAR-H fulfilled and surpassed the requirements for use in NA62, producing a photon yield 10-15\,\% greater with respect to the NA62 CEDAR Figure \ref{fig:Na62comp} and a kaon identification efficiency based on a five sector coincidence greater than 99.5\,\%.



\bibliographystyle{JHEP}
\bibliography{main.bib}

\end{document}